\documentclass[reprint,secnumarabic,amssymb, amsmath, nobibnotes, aps, prl]{revtex4-2}
\usepackage{lineno}
\usepackage{graphicx}
\usepackage{graphics}
\usepackage[colorlinks,
            linkcolor=blue,
            anchorcolor=blue,
            citecolor=blue
            ]{hyperref}
\usepackage{amsmath}
\usepackage{graphicx}
\usepackage{dcolumn}
\usepackage{bm}

\begin{document}

\title{Fano effect induced giant and robust enhancement of photon correlations in cavity QED systems}

\author{Yu-Wei Lu}%
\affiliation{School of Physics and Optoelectronic Engineering, Foshan University, Foshan 528000, China}
\author{Jing-Feng Liu}%
\affiliation{College of Electronic Engineering, South China Agricultural University, Guangzhou 510642, China.}
\author{Runhua Li}%
\affiliation{School of Physics and Optoelectronics, South China University of Technology, Guangzhou 510641, China}
\author{Haishu Tan}%
\email[Corresponding Author: ]{tanhaishu@fosu.edu.cn}
\affiliation{School of Physics and Optoelectronic Engineering, Foshan University, Foshan 528000, China}
\author{Yongyao Li}%
\affiliation{School of Physics and Optoelectronic Engineering, Foshan University, Foshan 528000, China}
\affiliation{Guangdong-Hong Kong-Macao Joint Laboratory for Intelligent Micro-Nano Optoelectronic
Technology, Foshan University, Foshan 528000, China}
\email[Corresponding Author: ]{yongyaoli@gmail.com}



\begin{abstract}
Fano effect arising from interference between two dissipation channels to the radiation continuum enables to tune the photon statistics. Understanding the role of Fano effect and exploiting to achieve strong photon correlations are of both fundamental and applied significance. We present an analytical description of Fano-enhanced photon correlations based on cavity quantum electrodynamics to show that, the Fano effect in atom-cavity systems can improve the degree of antibunching by over four orders of magnitude. The enhancement factors and the optimal conditions are explicitly given, and found related to the Fano parameter $q$. Remarkably, the Fano enhancement manifests robustness against the decoherence processes, and can survive in the weak coupling regime. We expect our work to provide insight in tuning the photon statistics through Fano effect, which offers a new route to enhance the photon correlations, as well as the possibility of generating nonclassical light in a wider diversity of systems without the need of strong light-matter interaction. 
\end{abstract}


\maketitle

Generation of nonclassical light is a key prerequisite for a number of quantum technologies, such as long-distance quantum communication \cite{RN1}, linear optical quantum computation \cite{RN2,RN3} and quantum cryptography \cite{RN5}. To generate single-photon, a conventional approach is found to utilize the quantum nonlinearities in energy levels for turning a coherent steam of photons (a laser field) into antibunched photons, giving rise to single-photon blockade (1PB) \cite{RN6,RN7,RN8,RN9}. 1PB can be implemented in different platforms, such as Kerr-nonlinearity cavities \cite{RN6} and optomechanical systems \cite{RN10}, but is typically achieved in strongly coupled atom-cavity systems with anharmonic Jaynes-Cummings ladder \cite{RN7,RN11,RN12}. The system enters the blockade regime when interactions overtake dissipation, and the photons start to antibunch. However, the effective blocking of second photon requires sufficiently strong interactions to increase the quantum nonlinearities.

In parallel, leakage is inevitable in all open optical cavities due to the coupling to radiation continuum. A sharp asymmetric lineshape appears when a localized resonant state of cavity interferes with the continuum of radiative states, known as the Fano resonance \cite{RN13,RN14}. Fano resonance has been widely observed in various nanophotonic structures involving the transmission and scattering of classical electromagnetic waves \cite{RN16,RN17}, and with important applications in optical sensing \cite{RN18,RN19} and switching \cite{RN21}. Since interferences are generally existed, the Fano effect also occurs in the quantum regime, especially in the emission process \cite{RN22,RN23,RN24,RN25} and the photon transport \cite{RN26,RN27}. Recently, several works have demonstrated the ability of Fano resonance to tune the photon statistics with a coherent input \cite{RN29,RN42,RN31}. Nevertheless, the current theories that tackles the Fano effect in photon statistics are dependent on the type of platforms \cite{RN29,RN42,RN30,RN31}, a more flexible and general theory based on cavity quantum electrodynamics (QED) is needed to explore the impact of Fano effect on photon statistics. Furthermore, it has been shown that the Fano resonance together with bound states in the continuum (BIC), an exotic kind of states that lie inside the radiation continuum but remain localized with no radiation \cite{RN32,RN33}, is critical in controlling over the photon statistics \cite{RN29}. The connection between the two phenomena, Fano effect and BIC, and their roles in enhancing the photon correlations, are ambiguous and need clarifying.  

In this Letter, we present a simple and flexible description of the Fano effect in tuning the photon correlations of cavity QED systems. The photon correlation function, the maximal enhancement due to the Fano effect and its analytical condition, are explicitly given in special parameters choice. Furthermore, we decompose the second-order correlation function following the squeezed-coherent Gaussian description \cite{RN8,RN34,RN35}, which is useful for analyzing how the Fano effect alters the quantum nature of interfering fields. 

\begin{figure}[ht]
\centering\includegraphics[width=\linewidth]{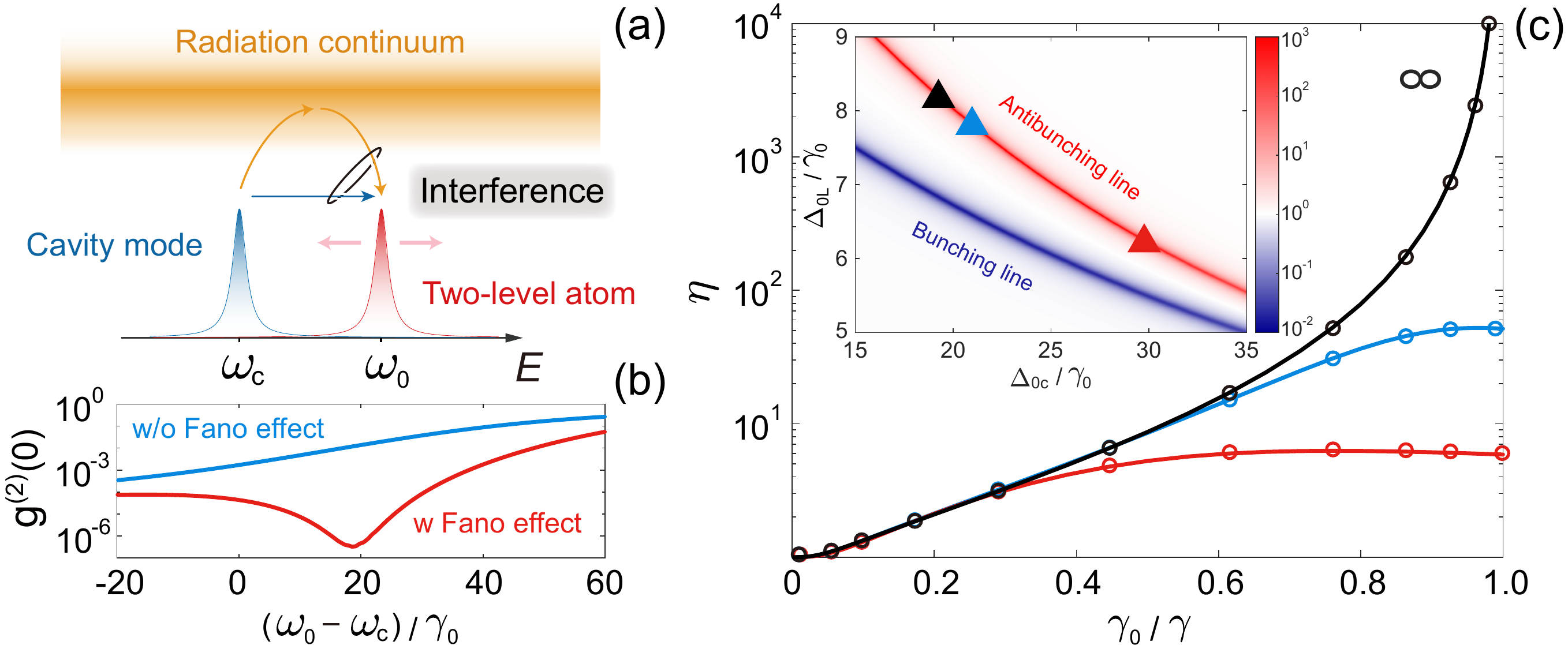}
\caption{(a) Schematic diagram of the Fano effect in a standard cavity QED system consisting of a two-level atom and a cavity mode. Two dissipation channels to the radiation continuum between atom and cavity interfere for the Fano effect. (b) The resulting second-order correlation function exhibits prominent enhancement over a wide range of atom-cavity detuning due to the occurrence of Fano effect. (c) The enhancement of photon correlations $\eta$ versus $\gamma_0/\gamma$ in cases of atom drive (solid lines) and cavity drive (hollow circles), for different detunings indicated by triangles with the same color in the inset, where $\eta$ is shown as the function of detunings $\Delta_{0L}$ and $\Delta_{0c}$.}
\label{fig1}
\end{figure}

Fig. \ref{fig1}(a) depicts the scenario of Fano interference in a standard Jaynes-Cummings (JC) system. A quantum emitter is modeled as two-level atom with transition frequency $\omega_0$, and coupled to a single-mode cavity with resonance frequency $\omega_c$. The atom and cavity interacting with a common radiation continuum represented by a Markovian bath, resulting in the radiative dissipation of each component and the interference for Fano effect. Taking into account the coherent driving and transforming into the rotating frame at driving frequency $\omega_L$, the total Hamiltonian of such a cavity QED system can be written as $H=H_0+H_C+H_B$, where the free Hamiltonian is $H_{0}=\Delta_{c L} c^{\dagger} c+\Delta_{0 L} \sigma_{+} \sigma_{-}+\sum_{\mu} \Delta_{\mu L} \alpha_{\mu}^{\dagger} \alpha_{\mu}$, with $\Delta_{XL}=\omega_X-\omega_L$ being the detuning for indices $X=c,0,\mu$. $\alpha_\mu$ is the bosonic annihilation (creation) operator for $\mu$-th radiative mode of in the common bath with frequency $\omega_\mu$. $\sigma_-$ and $c$ are the lowering operator and the bosonic annihilation operator of the atom and the cavity photon, respectively. The Hamiltonian of JC system driven by coherent fields is given by $H_{C}=g\left(c^{\dagger} \sigma_{-}+\sigma_{+} c\right)+\Omega_{0}\left(\sigma_{+}+\sigma_{-}\right)+\Omega_{c}\left(c^{\dagger}+c\right)$, where $\Omega_0$ ($\Omega_c$) is the amplitude of the driving field for atom (cavity). $H_{B}=\Sigma_{\mu} V_{\mu}\left(c^{\dagger} \alpha_{\mu}+\alpha_{\mu}^{\dagger} c\right)+\Sigma_{\mu} U_{\mu}\left(\sigma_{+} \alpha_{\mu}+\alpha_{\mu}^{\dagger} \sigma_{-}\right)$ describes interactions between the JC system and the radiation continuum, with $V_\mu$ and $U_\mu$ being the coupling constants of cavity-bath and atom-bath interactions for $\mu$-th radiative mode, respectively. 

The cavity-bath and atom-bath interactions follow the Markovian dynamics, and thus we can obtain the Heisenberg equations for cavity photon and atom by formally integrating and eliminating the equations of radiative modes
\begin{equation}
\dot{c}=-i\left(\Delta_{c L}-\frac{i \kappa}{2}\right) c-i\left(g-i \frac{\sqrt{\kappa_{0} \gamma_{0}}}{2}\right) \sigma_{-}+\Omega_{c}
\end{equation}
\begin{equation}
\dot{\sigma}_{-}=-i\left(\Delta_{0 L}-\frac{i \gamma}{2}\right) \sigma_{-}+i \sigma_{\mathrm{z}}\left(g-i \frac{\sqrt{\kappa_{0} \gamma_{0}}}{2}\right) a+\Omega_{0}
\end{equation}
where $\sigma_{\mathrm{z}}=\sigma_{+} \sigma_{-}-\sigma_{-} \sigma_{+}$. $\kappa=\kappa_0+\kappa_n$ and $\gamma=\gamma_0+\gamma_n$ are the total decay rates for photon and atom, respectively, with $\kappa_0=2\pi|V|^2$ and $\gamma_0=2\pi|U|^2$, where we have assumed the response of radiation continuum is flat enough so that the coupling constants are frequency independent and the subscripts have been dropped. $\kappa_n$ and $\gamma_n$ represent the additional non-radiative and radiative dissipation due to the interactions of cavity and atom with the independent baths, respectively, which are introduced in a similar manner as the coupling to the radiation continuum. The above equations of motion are equivalent to a non-Hermitian effective Hamiltonian
\begin{equation}
\begin{aligned}
H_{\text {eff }}=& \Delta_{c} c^{\dagger} c+\Delta_{0} \sigma_{+} \sigma_{-}+g_{F}\left(c^{\dagger} \sigma_{-}+\sigma_{+} c\right) \\
&+\Omega_{0}\left(\sigma_{+}+\sigma_{-}\right)+\Omega_{c}\left(c^{\dagger}+c\right)
\end{aligned}
\end{equation}
with notations $\Delta_c=\Delta_{cL}-i\kappa/2$, $\Delta_0=\Delta_{0L}-i\gamma/2$ and $g_{F}=g-i \sqrt{\kappa_{0} \gamma_{0}} / 2$. In the low-driving regime, the pump field only perturbs the ground state system. By truncating the state space at two-excitation manifold, the intensity $n_{c}=\left\langle c^{\dagger} c\right\rangle$ and the second-order correlation function $g^{(2)}(0)=\left\langle c^{\dagger^{2}} c^{2}\right\rangle / n_{c}$ of cavity photon can be analytical obtained through the wavefunction approximation with the effective Hamiltonian $H_\mathbf{eff}$ \cite{RN34,RN36,RN46}, which are expressed as
\begin{equation}\label{eq4}
n_{c}=\left|\frac{\Omega_{c} \Delta_{0}-\Omega_{0} g_{F}}{g_{F}^{2}-\Delta_{0} \Delta_{c}}\right|^{2}
\end{equation}
\begin{equation}\label{eq5}
g^{(2)}(0)=\left|\frac{g_{F}^{2} \Omega_{0}^{2}-2 g_{F}\left(\Delta_{0}+\Delta_{c}\right) \Omega_{0}+\Omega_{c}^{2}\left[g_{F}^{2}+\Delta_{0}\left(\Delta_{0}+\Delta_{c}\right)\right]}{\left(g_{F}^{2}-\Delta_{0} \Delta_{c}\right)\left[g_{F}^{2}-\left(\Delta_{0}+\Delta_{c}\right) \Delta_{c}\right]}\right|^{2}
\end{equation}
The obtained Eqs. (\ref{eq4})-(\ref{eq5}) capture the Fano effect in cavity QED systems, and we define the enhancement of photon correlations as $\eta=g_{0}^{(2)}(0) / g^{(2)}(0)$, where $g_{0}^{(2)}(0)$ is the second-order correlation function without the Fano effect. Fig. \ref{fig1}(b) compares the minimal $g^{(2)}(0)$ with and without the Fano effect for atom drive by varying $\omega_0$, with parameters $g=15\gamma_0$, $\gamma_n=10^{-2} \gamma_0$, $\kappa_0=0.3\gamma_0$ and $\kappa_n=0$. It shows that the Fano effect greatly enhances the photon correlations by over four orders of magnitude, achieved with detuned atom-cavity coupling. The corresponding detuning $\left(\Delta_{0 c}, \Delta_{0 L}\right)=(19.25 \gamma_{0},8.2 \gamma_{0})$ is indicated by the black triangle in the inset of Fig. \ref{fig1}(c), where $\Delta_{0 c} = \omega_0 - \omega_c$. At this special point, we find that the enhancement of photon correlations $\eta$ is diverging as $\gamma_n$ decreases, see the black line in Fig. \ref{fig1}(c); by contract, the maximal $\eta$ remains finite for other parameters (blue and red lines). Though the diverging $\eta$ only exists with the two-excitation truncation, this feature implies the formation of BIC in the system, since Eq. (\ref{eq5}) indicates that $\eta$ diverges due to $\left(g_{F}^{2}-\Delta_{0} \Delta_{c}\right) \rightarrow 0$ when $\gamma_{n}, \kappa_{n} \rightarrow 0$, which is exactly the mechanism of Friedrich–Wintgen BIC \cite{RN32,RN33}. This can be verified by finding the optimal $\Delta_{0c}$ corresponding to the minimum decay of one of the eigenenergies in the single-excitation subspace of $H_\mathbf{eff}$, which is given by
\begin{equation}\label{eq6}
\Delta_{0 c}^{B I C} \approx q\left[\left(1-\beta_{\kappa}\right) \gamma_{0}-\left(1-\beta_{\gamma}\right) \kappa_{0}\right] / 2
\end{equation}
where $\beta_\kappa=\kappa_n/\kappa$ and $\beta_\gamma=\gamma_n/\gamma$ are the ratios specifying the fraction of dissipation into the independent baths other than the common radiation continuum for cavity and atom, respectively, and we have expanded the optimal $\Delta_{0c}$ up to first order with respect to $\beta_\kappa$ and $\beta_\gamma$. $q=2 g / \sqrt{\kappa_{0} \gamma_{0}}$ is the Fano parameter. We can see that Eq. (\ref{eq6}) becomes the condition of Friedrich–Wintgen BIC if we take $\gamma_n=\kappa_n=0$ \cite{RN32,RN33,RN37}. Therefore, BIC is a special regime of the Fano effect that can generate nonclassical light with much stronger correlations. However, in general $\gamma_n, \kappa_n \neq 0$, the system is under the quasi-BIC regime with a finite $\eta$. The maximal $\eta$ is found located at
\begin{equation}
\Delta_{0 L}^{B I C} \approx q\left(1-\beta_{\gamma}\right) \kappa_{0} / 2
\end{equation}

\begin{figure}[t]
\centering\includegraphics[width=\linewidth]{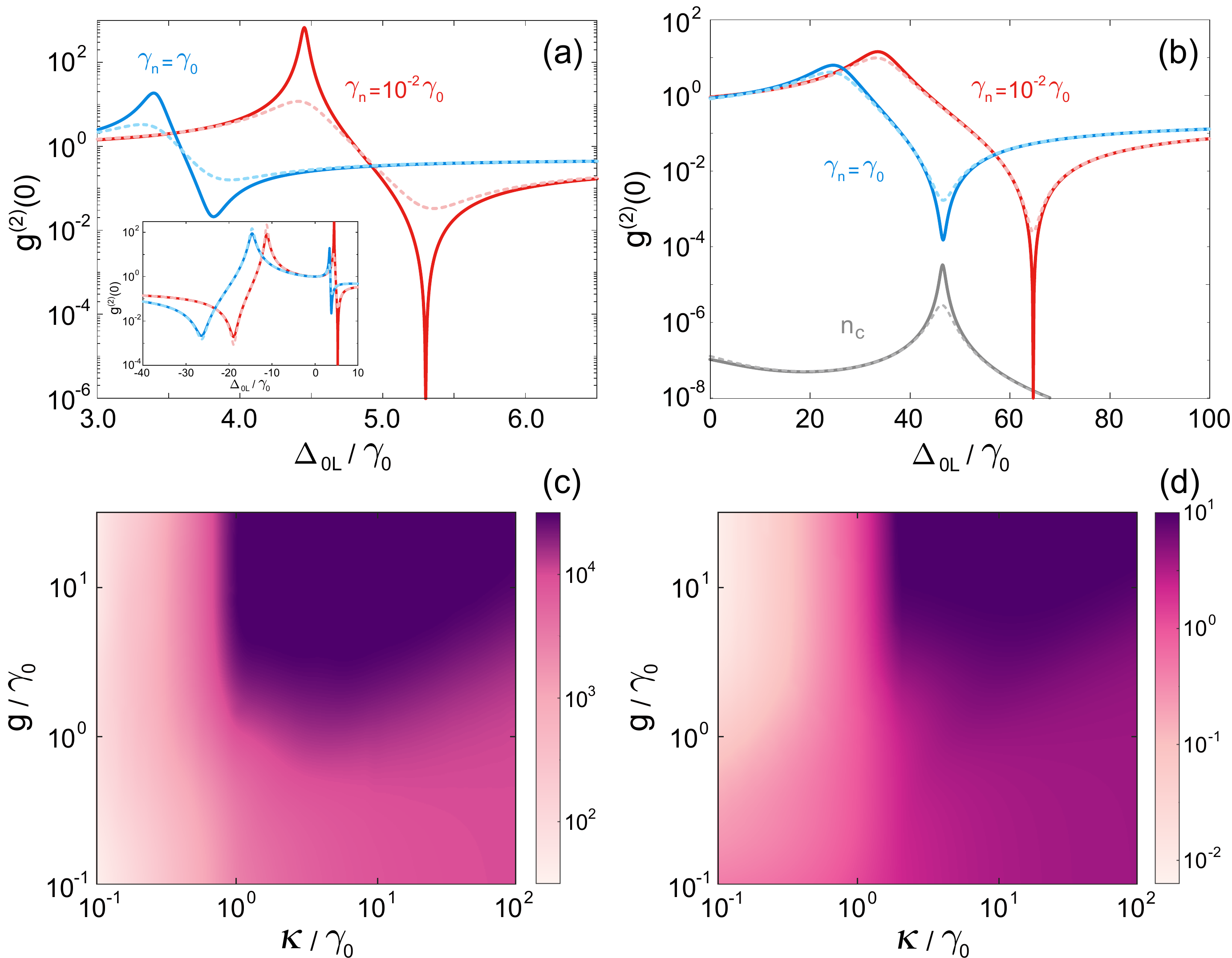}
\caption{(a) and (b) Comparison of photon correlations with (solid lines) and without (dashed lines) the Fano effect for atom drive with parameters $\left(g, \kappa_{0}\right)=\left(\gamma_{0}, 0.3 \gamma_{0}\right)$ and $\left(g, \kappa_{0}\right)=\left(16 \gamma_{0}, 17 \gamma_{0}\right)$. (c) and (d) The ratio of minimum $g^{(2)}(0)$ to $g_F^{(2)}(0)$ as the function of $g$ and $\kappa$ for $\gamma_n=10^{-2} \gamma_0$ and $\gamma_n=\gamma_0$, respectively. $\kappa_n=0$ for all figures.}
\label{fig2}
\end{figure}

As Eq. (\ref{eq4}) indicates, $\Delta_{0 L}^{B I C}$ is corresponding to the Fano maximum in emission spectrum, see Fig. \ref{fig2}(b) for an example. Accordingly, the maximal enhancement of photon correlations $\eta_m$ can be analytically obtained with $\Delta_{0 c}^{B I C}$ and $\Delta_{0 L}^{B I C}$, and the corresponding $g^{(2)}(0)$ is denoted as $g_F^{(2)}(0)$ hereafter. We are interesting in two limiting cases, $q \approx 0$ and $q \gg 1 $. For the former case, we have
\begin{equation}\label{eq8}
\eta_{m} \approx\left[1+\frac{1}{\left(1+\beta_{\gamma}\right)\left(1+\beta_{\kappa}\right)-1}\right]^{2}\left(\frac{\kappa_{0}}{C_{X} \kappa_{0}+\gamma_{0}}\right)^{2}
\end{equation}
where the coefficient $C_X=1$ for $X=0$ (atom drive) and $C_X=2$ for $X=c$ (cavity drive). While for $q \gg 1$, we obtain
\begin{equation}\label{eq9}
\eta_{m} \approx \frac{\left(2+\beta_{\gamma}+\beta_{\kappa}\right)^{2} q^{2}+\left(1+\beta_{\gamma}\right)^{2}\left(1+\beta_{\kappa}\right)^{2}}{\left(\beta_{\gamma}+\beta_{\kappa}\right)^{2} q^{2}+\left[\left(1+\beta_{\gamma}\right)\left(1+\beta_{\kappa}\right)-1\right]^{2}} D_{X}
\end{equation}
where the coefficient $D_{X}=1-4\left(\kappa_{0}+\gamma_{0}\right) \gamma_{0} /\left(q \kappa_{0}\right)^{2}$ for atom drive and $D_{X}=1-4\left(\kappa_{0}+\gamma_{0}\right)^{2} /\left(q \kappa_{0}\right)^{2}$ for cavity drive. The results show that, though the Fano effect can achieve stronger correlations for both types of driving, the enhancement factors are slightly different. From the coefficients $C_X$ and $D_X$, we can see that the enhancement of photon correlations under atom drive is greater than cavity drive. Therefore, driving the atom can obtain single photon with higher purity, since $g^{(2)}(0)$ at $\Delta_{0 L}^{B I C}$ is also smaller in this case.

Eqs. (\ref{eq8}) and (\ref{eq9}) indicate that the additional dissipation degrades the Fano enhancement of photon correlations. Fig. \ref{fig2}(a) displays that, despite $\eta_m$ declines by over three orders of magnitude as $\gamma_n$ increases from $10^{-2} \gamma_0$ to $\gamma_0$, the Fano effect still produces a tenfold enhancement of photon correlations with $g / g_{c} \sim 1.5$, where $g_c=(\kappa+\gamma)/2$ is the critical coupling constant for strong coupling \cite{RN9}. While for a system with greater dissipation but similar $g/g_c$, $\eta_m>10$ is also observed, as Fig. \ref{fig2}(b) shows. It demonstrates the robustness of Fano enhancement against the decoherence. Besides, it also shows that the Fano effect can lead to stronger bunching of photons, and thus offers great tunbility to photon statistics. Note that though the maximal reduction of $g^{(2)}(0)$ occurs at $\Delta_{0 L}^{B I C}$, the minimum $g^{(2)}(0)$ might not be $g_F^{(2)}(0)$, see the inset of Fig. \ref{fig2}(a) for an example, where the minimum achieves at another $g^{(2)}(0)$ dip. Therefore, in Fig. \ref{fig2}(c) and (d) we show the ratio of minimum $g^{(2)}(0)$ to $g_F^{(2)}(0)$ versus $g$ and $\kappa$ for $\gamma_n=10^{-2} \gamma_0$ and $\gamma_n=\gamma_0$, respectively. We can clearly see that the Fano enhancement has significant advantage with increasing $\kappa_0$, even greater dissipation is introduced in the system. It is because the increase of $\kappa_0$ simultaneously reduces $\beta_\kappa$, resulting in stronger Fano effect, and thus inverses the detrimental effect of dissipation on photon correlations. This feature is more obvious for $g>10\gamma_0$, where the photon correlations exhibits an abrupt enhancement around $\kappa_0=\gamma_0$. However, a larger $\kappa_0$ is not always beneficial though the Fano enhancement can persist. As Fig. \ref{fig2}(c) and (d) show, there is an optimal $\kappa_0$ of $5\sim 10\gamma_0$ for the maximal enhancement of photon correlations, as a consequence of balance between the negative impact of dissipation and the enhancement from Fano effect. 

\begin{figure}[t]
\centering\includegraphics[width=\linewidth]{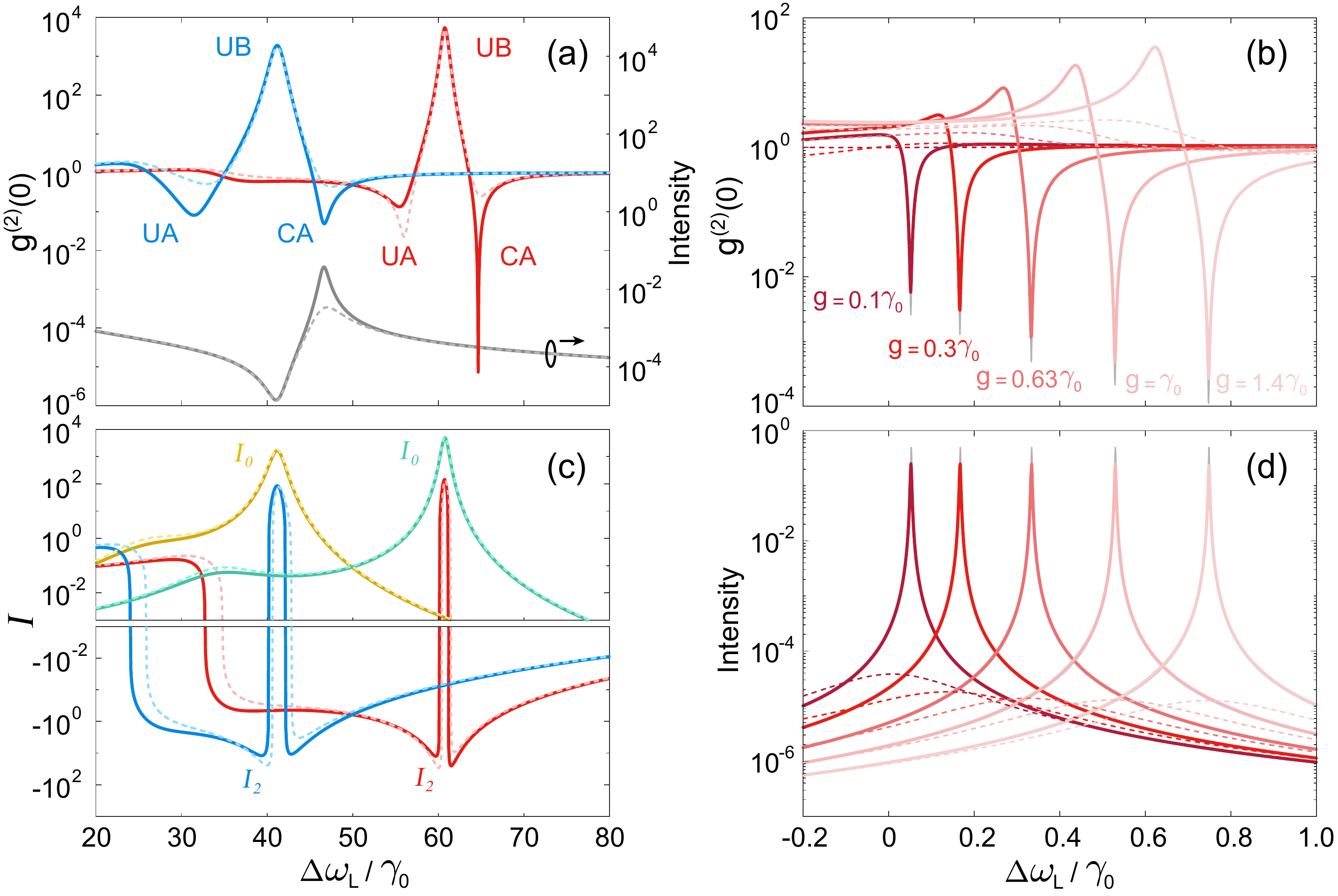}
\caption{(a) and (c) Comparison of photon correlations and corresponding decomposition with (solid lines) and without (dashed lines) the Fano effect for cavity drive. The parameters are the same as Fig. 2(b), and $\Omega_c=\gamma_0$. (b) and (d) Comparison of photon correlations and intensity with (solid lines) and without (dashed lines) the Fano effect for cavity drive and various $g$. The parameters are $\left(\gamma_{n}, \kappa_{0}\right)=\left(10^{-2} \gamma_{0}, 0.3 \gamma_{0}\right)$. The numerical results are obtained using QuTip \cite{RN23,RN39} with $\Omega_c=10^{-3} \gamma_0$, and the results of analytical expressions Eqs. (\ref{eq4}) and (\ref{eq5}) are shown for comparison (light gray lines). }
\label{fig3}
\end{figure}

For cavity driven case, the characteristics of $g^{(2)}(0)$ are more complex than the atom drive, due to the existence of a new mechanism for photon antibunching, which originates from the quantum interference between different transition pathways, i.e., so-called unconventional antibunching (UA) \cite{RN8,RN40,RN41}, and distinguishes from the conventional antibunching (CA) taking place at the Fano maximum. As Fig. \ref{fig3}(a) shows, an additional $g^{(2)}(0)$ dip emerges at the left side of bunching peak. We can see from Fig. \ref{fig3}(a) that the Fano effect can enhance not only CA, but also UA for larger $\gamma_n$ (blue lines). 

To better understand how the Fano effect modifies the quantum nature of light for different circumstances, we decompose $g^{(2)}(0)$ of JC system according to the picture of squeezing-coherent state admixture \cite{RN34,RN35}: $g^{(2)}(0)=1+I_0+I_2$, where $I_{0}=\left(\left\langle c^{\dagger 2} c^{2}\right\rangle+\left|\left\langle c \right\rangle\right|^4-2 \operatorname{Re}\left[\left\langle c^{\dagger}\right\rangle^{2}\left\langle c^{2}\right\rangle\right]\right) / n_{c}^{2}$ and $I_{2}=2 \operatorname{Re}\left(\left[\left\langle c^{\dagger}\right\rangle^{2}\left\langle c^{2}\right\rangle\right]-\left|\left\langle c \right\rangle\right|^4\right) / n_{c}^{2}$ represent the sub-Poissonian statistics and the squeezing of light, respectively, and thus the decomposition describes the emitted light as the mixing of a quantum signal of atom and a coherent field from cavity. The noncoherent contribution $I_0$ reads as
\begin{equation}
I_{0}=\left|\frac{g_{F}^{4}}{\Delta_{0}^{2}\left[g_{F}^{2}-\left(\Delta_{0}+\Delta_{c}\right) \Delta_{c}\right]}\right|^{2}
\end{equation}
$I_2$ can be subsequently evaluated as $I_2=g^{(2)}(0)-1-I_0$. Fig. \ref{fig3}(c) compares the decomposition of $g^{(2)}(0)$ with and without the Fano effect, where we can clearly see that for $\gamma_n=\gamma_0$, the decrease of $I_2$ is responsible for the Fano-enhanced photon correlations around UA. On the contrary, $I_2$ of $\gamma_n=10^{-2} \gamma_0$ increases at UA, leading to the opposite effect. Furthermore, we notice that both $I_0$ and $I_2$ around CA are steeper after introducing the Fano effect, resulting in the enhancement of photon correlations. Therefore, the decomposition of $g^{(2)}(0)$ reveals that the Fano-enhanced photon correlations is mainly contributed by the increased squeezing component of emitted light.  

Fig. \ref{fig3}(b) and (d) plot $g^{(2)}(0)$ and intensity for various $g$, where entering the strong coupling regime requires $g>g_c=0.65\gamma_0$. We can see that strongly antibunched light with $g^{(2)}(0) \sim 10^{-3}$ is observed below the strong coupling, and $g^{(2)}(0)<10^{-2}$ can be achieved even with $g=0.1\gamma_0$ due to the Fano enhancement, contrast to the flat photon correlations with $g^{(2)}(0) \sim 1$ in absence of Fano effect. Moreover, the Fano effect also gives rise to a sharp resonance in the cavity emission spectrum, accompanied by a giant enhancement ($\sim 10^4$) of intensity, as Fig. \ref{fig3}(d) shows. The population of cavity photon reaches $\sim 0.3$ and steadies as $g$ decreases from $1.4\gamma_0$ to $0.1\gamma_0$. The results demonstrates the peculiarity of Fano-enhanced photon correlations that can simultaneously realize strong antibunching and high intensity, which is attractive for practical applications. 

However, though generally existed, the Fano effect is often too weak to have significant impact on the photon statistics, since in most cases it is induced by the free-space continuum. To enhance the Fano effect, one can couple the JC system to a waveguide, which takes over the role of radiation continuum. More importantly, $\beta_\kappa$ and $\beta_\gamma$ are related to the decay rates of cavity and atom to the guided mode, respectively, and thus can be feasibly controlled by engineering the evanescent coupling between the JC system and the waveguide. 

To summarize, we present an analytical description of Fano-enhanced photon correlations for cavity QED, and show its connection with Friedrich–Wintgen BIC. The Fano effect manifests considerable enhancement of photon correlations, achieved by increasing the squeezing of light. The Fano enhancement can persist with large dissipation, and can produce strongly antibunched light with high population even in the weak coupling regime. Therefore, the Fano-enhanced photon antibunching holds great potential in realizing practical single-photon source in a variety of platforms, including PC waveguide \cite{RN29,RN42}, hybrid plasmonic-photonic cavity \cite{RN43,RN45}, and open cavity magnonic system \cite{RN47}.

\bibliography{FPB.bib}

\end{document}